\documentclass[ prd,twocolumn,superscriptaddress]{revtex4-2}
\usepackage{geometry}                % See geometry.pdf to learn the layout options. There are lots.
\usepackage{graphicx}
\usepackage{amssymb}
\usepackage{amsmath}
\usepackage{epstopdf}
\usepackage{xspace}
\usepackage[dvipsnames]{xcolor}
\usepackage{breqn}
\newcommand{\gf}{G\textsc{eant}4\xspace}

\graphicspath{ {./} }

\begin{document}

\title{Simulating the charging of isolated free-falling masses from TeV to eV energies: detailed comparison with LISA Pathfinder results}

\def\addressd{High Energy Physics Group, Physics Department, Imperial College London, Blackett Laboratory, Prince Consort Road, London, SW7 2BW, UK }
\def\addressbb{Department of Mechanical and Aerospace Engineering, MAE-A, P.O. Box 116250, University of Florida, Gainesville, Florida 32611, USA}
\def\addressff{Department of Physics, P.O. Box 118440, University of Florida, Gainesville, Florida 32611, USA}

\author{P.\,J.~Wass}\thanks{Corresponding author}\affiliation{\addressbb}\affiliation{\addressd}
\author{T.\,J.~Sumner}\affiliation{\addressd}\affiliation{\addressff}
\author{H.\,M.~Ara\'{u}jo}\affiliation{\addressd}
\author{D.~Hollington}\affiliation{\addressd}

\date{draft of \today}                                           % Activate to display a given date or no date

%%%%%%%%%%-----------------------------------------------------------------------%%%%%%%%%
%%%%%%%%%%-----------------------------------------------------------------------%%%%%%%%%

\begin{abstract}
A model is presented that explains the charging rate of the LISA Pathfinder test masses by the interplanetary cosmic ray environment. The model incorporates particle-tracking from TeV to eV energies using a combination of \gf and a custom low-energy particle generation and tracking code. The electrostatic environment of the test mass is simulated allowing for a comparison of the test-mass charging-rate dependence on local electric fields with observations made in orbit. The model is able to reproduce the observed charging behavior with good accuracy using gold surface properties compatible with literature values. 
The results of the model confirm that a significant fraction of the net charging current is caused by a population of low-energy ($\sim$eV) electrons produced by electron- and ion-induced kinetic emission from the test mass and surrounding metal surfaces. Assuming a gold work function of 4.2\,eV, the unbalanced flow of these electrons to and from the unbiased test mass contributes $\sim$10\% of the overall test mass charging rate. Their contribution to the charging-current shot noise is disproportionately higher and it adds $\sim$40\% to the overall predicted noise.  However, even with this increased noise contribution the overall charging-current noise is still only 40\% of that measured in-orbit, and this remains an unsolved issue.

\end{abstract}
\maketitle

%%%%%%%%%%-----------------------------------------------------------------------%%%%%%%%%
%%%%%%%%%%-----------------------------------------------------------------------%%%%%%%%%

\section{Introduction}
\label{intro}

LISA Pathfinder (LPF)~\cite{McNamara2008} was a European Space Agency mission to test the performance of free-falling test masses (TM) for the future gravitational wave observatory the Laser Interferometer Space Antenna (LISA)~\cite{LISA2017}. Electrostatic charge build up on the test masses can compromise the performance of LISA through the interaction with stray electric fields~\cite{jafry1996}. Test mass charging due to the space environment has been measured during the LPF mission~\cite{Armano2017a,Armano2022} but understanding the physical processes that lead to the observed charging rates is of importance for the design of future space missions which may have different geometries or operate in different environments.

The TM charge accumulates due to the flux of energetic charged particles in the space environment. Cosmic rays with energy above $\sim$100\,MeV/nucleon are able to penetrate the shielding of the spacecraft and deposit charge on the TM~\cite{jafry1996}. The incoming spectrum extends many orders of magnitude above 100\,MeV/nucleon making the problem suitable for simulation by high-energy physics simulation tools.
During the development phase of the LISA Pathfinder mission, a number of studies simulated test mass charging due to galactic cosmic rays and solar energetic particles in LISA and LPF~\cite{Araujo2005,Wass2005,Grimani2005,Grimani2005b, Grimani2015} using \gf~\cite{geant4} and FLUKA~\cite{fluka} high-energy physics simulation toolkits. This paper updates and extends the previous modelling using \gf to include a detailed assessment of the low energy electron population needed to explain new LPF results which show a dependency of charging rate on the TM potential~\cite{Armano2022}.

In the remainder of this introduction we summarize the results of previous work on modelling TM charging and introduce the relevant features of the LPF instrument.  In Sec.~\ref{model:charging} the approach taken for the modelling of the charging process, including the basic particle transport and charge transfer, is described. In addition, a simple analytical toy model is presented to supplement the Monte Carlo method to illustrate the effect of the low-energy electron population.   
Sec.~\ref{results:Q_rate} presents the results from the both the Monte Carlo model and the toy model, and uses these to confront the LPF measurements.  A discussion of the uncertainties of the charging modelling is given in Sec.~\ref{discussion}, and in Sec.~\ref{Discussion2} the conclusions of the work are presented.

%%%%%%%%%%-----------------------------------------------------------------------%%%%%%%%%

Predictions of the mean charging rate from modelling processes above $\sim$250\,eV presented in \cite{Araujo2005, Wass2005} were between +20 and +25\,e\,s$^{-1}$ (where e is the proton charge) at solar maximum---and double that at solar minimum. The statistical nature of the charge build up was investigated and predictions made for the associated charge noise behaviour. 
Whilst much of the charging was due to simple proton stopping, there were individual net charge deposits as high(low) as +80($-$60)\,e, with some events involving a few hundred charges at a time with only a fraction remaining within the TM. Thus, the mean charging rate is the result of a large number of events producing both negative and positive charging and with some higher multiplicity events of both polarities. Hence, the simulated stochastic noise behaviour in the charging rate was higher than expected from the mean rate and could be characterised by an effective single charge current of $\lambda_{\mathrm{eff}}\sim$200--400\,e\,s$^{-1}$. The observed charging rate noise in orbit was significantly higher~\cite{Armano2017a}, and this paper will also address this, but without satisfactory resolution at the moment.

\gf was not developed to model processes at eV energies. Instead, the charging contribution from these particles was estimated from the fluxes of higher energy particles crossing the TM and surrounding boundaries, and assessing their secondary-particle production yields for a variety of low-energy processes. This suggested that the mean charging rate could be significantly changed by such a low-energy population and, moreover, that it might show a dependence on applied $\sim$eV potentials. 

The gravitational reference sensor (GRS)~\cite{Dolesi2003} on LPF comprised a $\sim$2\,kg gold-plated Au/Pt(70/30) TM within a six-degree-of-freedom capacitive position sensor and actuator. The TM had no electrical connection to the ground of the spacecraft. 12 gold-plated electrodes separated from the TM by mm gaps were used for position sensing and actuation~\cite{Armano2017b}. Voltages with amplitudes of 0.5-10\,V were applied for force and torque actuation at ac audio frequencies, while dc or slowly varying voltages could also be applied as needed. A further six electrodes on the $y$ and $z$ were used to induce a 0.6\,V TM potential at 100\,kHz, used for position sensing. The other applied ac voltages were applied in a manner that did not produce a net polarization of the TM.

%%%%%%%%%%-----------------------------------------------------------------------%%%%%%%%%
%%%%%%%%%%-----------------------------------------------------------------------%%%%%%%%%

\section{Charging predictions}
\label{model:charging}

%%%%%%%%%%-----------------------------------------------------------------------%%%%%%%%%

In this section the \gf charging prediction models used for LISA and LPF are described, taking advantage of number of improvements outlined in Sec.~\ref{model:G4_sims}, and building in new constraints from the measurements of charging rate dependence on potential presented here for the first time. In particular, a more extensive model of the production of low-energy secondary electrons via electron-induced electron emission (EIEE) and ion-induced electron emission (IIEE) (which are collectively referred to as kinetic emission (KE)) is developed, along with enhancements to the simulation code to follow the charge transfer processes within the GRS itself~\cite{Hollington2011}. The extended model is described in Sec.~\ref{model:KE_process}. 
There has been other recent relevant work on secondary electron emission applied to LPF~\cite{grimani21a, grimani21b, grimani22,taioli22} in order to explain the higher than expected charging rate noise levels observed, but which does not address the new constraints arising from the dependence of charging rate on potential.

%%%%%%%%%%-----------------------------------------------------------------------%%%%%%%%%

\subsection{Energetic particle interactions}
\label{model:G4_sims}

In studying the interaction of the primary cosmic ray particles with the spacecraft, the simulations used in \cite{Araujo2005} are revisited, implementing a number of updates. Firstly the geometry of the LPF spacecraft has been updated. While a full reproduction of the as-flown spacecraft was deemed unnecessarily complex, a number of changes were made to the implementation of the payload elements closest to the TM to reflect design changes that occurred late in the development of the mission. A number of dimensions of electrodes within the housing have been adjusted, the material of the GRS electrodes has been changed from sapphire to molybdenum and the tungsten gravitational balance masses surrounding the GRS were adjusted to match the size and location implemented in the final design of the instrument. As with previous simulations, the total mass simulated in the spacecraft is approximately 80\% of the total in-orbit mass of LPF.

Secondly, simulations have been executed using \gf version 10.3, which has had significant development of the software toolkit since the publication of the previous work~\cite{Araujo2005} which used \gf version 6.0. Of key importance to the test-mass charging process are the hadronic models that deal with interactions of primary cosmic rays and their secondaries with energies above 100\,MeV/nucleon, and electromagnetic processes that govern the production of low-energy 
particles, close to the TM. 

The physics models used by the \gf simulations and the energies over which they are applied are defined by the simulation physics list. A comprehensive overview of the physics models in \gf can be found in the physics reference manual for release 10.3 used in this work~\footnote{http://cern.ch/geant4-userdoc/UsersGuides/ForApplicationDeveloper/ BackupVersions/V10.3/fo/BookForAppliDev.pdf}.
Highlighted below are changes relevant to the high and low energy regimes relative to the previous implementation in Ref.~\cite{Araujo2005}.

The baseline simulations use the \texttt{FTFP\_Bert} reference physics list for hadronic interactions which is most commonly used for high-energy and space applications \cite{allison2016}. The \texttt{QGSP_BiC} physisc list was also used to investigate the effect of these models on the net charging rate.

Low-energy charged particle interactions including the photoelectric effect, Compton scattering, Rayleigh scattering, gamma conversion, bremsstrahlung and ionization are governed by electromagnetic (EM) physics models. These processes are well understood and validated but careful choices still need to be made in striking the balance between computational efficiency and accuracy. Any charged particle can contribute to charging irrespective of its energy and so the priority is accuracy of the simulation over computational cost, especially in the production of low-energy secondary particles. 
All particles created in the simulation are tracked to zero energy, but the minimum energy for secondary particle production (especially delta electrons from ionisation) depends on the EM physics model and on the incident particle type. The earlier work \cite{Araujo2005} used a customised physics list with a 250\,eV minimum production cut for electrons.  For the Livermore EM physics list used in this work, the recommended minimum for secondary ionisation produced by primary electrons is also
around 250\,eV, although the coverage of some models can be extended by extrapolation to as low as 10\,eV, (see for example Ref.~\cite{sakata2016}).   This work limits the minimum energy of electrons to 100\,eV to avoid compromising the accuracy of the simulation.  For ionisation produced by primary protons, the minimum production cut in the Livermore EM physics is at 790\,eV, the mean ionisation energy for gold. The uncertainties resulting from these production cuts is discussed in Sec.~\ref{discussion}.  
Atomic deexcitation processes were set to default (with fluorescence enabled, but Auger emission, Auger cascades and PIXE not enabled) for the full simulation model. However, their full contributions in terms of additional electron production were checked by a simpler simulation. In the case of 1\,MeV electron primaries for example there is a small increase at the few percent level.  For 300\,MeV protons there was no detectable increase seen in a simulation with 10$^6$ primaries.
The Livermore EM physics list makes use of the \texttt{G4UrbanMSCoption4} physics model for multiple scattering physics which provides the most accurate representation of scattering at boundaries.

%%%%%%%%%%-----------------------------------------------------------------------%%%%%%%%%

\subsection{Kinetic emission processes}
\label{model:KE_process}

Several low-energy physical processes below 100\,eV are not modelled by \gf but have the potential to significantly change the results; these were identified in the original work~\cite{Araujo2005}. In particular, low-energy electrons can be ejected when a higher energy charged particle crosses a surface boundary.  These kinetic electrons can be ejected by either electrons (EIEE) or ions (IIEE). The KE will come from the surface layers of the TM and EH and these are all gold plated with sufficient thickness that contains the process to that layer. 

{\it KE yield.}
The KE yield is energy dependent and Ref.~\cite{Araujo2005} used several literature measurements to estimate the full curves for gold from electrons, protons and alpha particles at normal incidence, to make a preliminary estimate of contribution to TM charging. This work parameterizes the energy and angular dependence of the KE yield to calculate the KE emission probability particle by particle. 

A number of sources in the literature employ semi-empirical models to parameterize the KE yield 
as a function of angle and energy due to electrons \cite{ludwick2020,furman2002,bundaleski2015} and ions~\cite{hasselkemp1992} incident on gold. 

\begin{figure}[htb]
\centering
\includegraphics[scale=0.48, trim=3cm 3cm 4cm 14cm, clip]{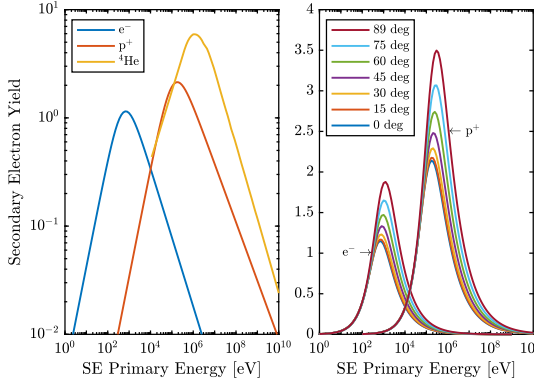}
\caption[KE Yield Curves.]{\label{fig:KEyieldCurves} The kinetic electron yield curves for gold with incident electrons, protons and alpha particles.  The left-hand panel is for normal incidence using Eq.~\ref{ke_ang} with $s=1.64$, $Y_0$=1.15 and $E_0=700$\,eV.  The right-hand panel shows the dependence on angle of incidence according to Eq.~\ref{ke_ang} as implemented in the simulations for electrons (left population) and protons(right population).}
\end{figure}

Preliminary calculations using several of these models with an angular dependence close to $\cos^{-1}\theta$ produced an average yield with a very large number of KEs in strong disagreement with TM charging noise results \cite{Armano2016} and with the new measurements of the dependence of charging rate on $V_{\textsc{tm}}$ presented in Ref.~\cite{Armano2022}. Very high yield at grazing incidence can be avoided by decreasing the negative exponent of the cosine dependence. It was found that the model of Furman and Pivi~\cite{furman2002} provided a reasonable match to the experimental results. They describe the KE yield, $Y$, due to an electron with energy $E$ incident at an angle relative the surface normal $\theta$ as
\begin{equation}
	\label{ke_ang} 
		Y(E, \theta)=Y_{\mathrm{max}}(\theta) \frac{ s E_r(\theta)}{s-1+E_r(\theta)^s},
\end{equation}
where $E_r = E/E_{\mathrm{max}}(\theta)$, with
\begin{equation}
		E_{\mathrm{max}}(\theta) =  E_0 \left[ 1 + 0.7(1-\cos{\theta})\right],
\end{equation}
and 
\begin{equation}
		Y_{\mathrm{max}}(\theta)= Y_0 \left[1 + 0.66(1-\cos^{0.8}{\theta})\right].
\end{equation}
$Y_0$ and $E_0$ are the values of yield and peak energy, respectively, for particles normally incident on the surface. A number of values for $Y_0$ in EIEE, can be found in the literature ranging from 1.28 to 2.5~\cite{lin2005,lundgreen2020,azzolini2019} and thus this was treated as a somewhat free parameter to be constrained by the experimental data. Indeed, the actual yield realized for real surfaces is known to depend on its condition and contamination state~\cite{gonzalez2017,yang2016}.
$E_0$ defines the peak of the yield curve measured between 500 and 800\,eV~\cite{lin2005,lundgreen2020,azzolini2019} for electrons. 
The parameter $s$ describes the spectral index of the yield curve for $E_r > E_{\mathrm{max}}$. Literature values range from 1.5 to 1.7 \cite{furman2002,lin2005,lundgreen2020}. The effect on the overall yield of varying both $E_0$ and $s$ is degenerate with $Y_0$ and for this work we choose to fix values of $s=1.64$ and $E_0=700$\,eV for EIEE as these parameters are relatively well constrained by the literature. A quantitative discussion of the impact of these parameters on the agreement between our simulations and measured data is given in Section~\ref{discussion}.
The right-hand panel in Fig.~\ref{fig:KEyieldCurves} illustrates how the yield curves for incident primary electrons and protons vary with angle of incidence. For IIEE due to protons, which is the subdominant process in our simulation we fix $Y_0=2.18$, $E_0=180$\,keV and $s$=1.54 \cite{haque2019}. Note that $E_0$ scales approximately with the energy of maximum stopping power for the corresponding particle. IIEE due to alpha particles is small and the angular dependence of the yield is not considered. 
In principle the emission properties for EIEE and IIEE may vary from surface to surface within the GRS, however, in order not to introduce too many parameters to the model, the emission properties of all surfaces are assumed identical. 

\begin{figure}[htb]
\centering
\includegraphics[width=0.45\textwidth]{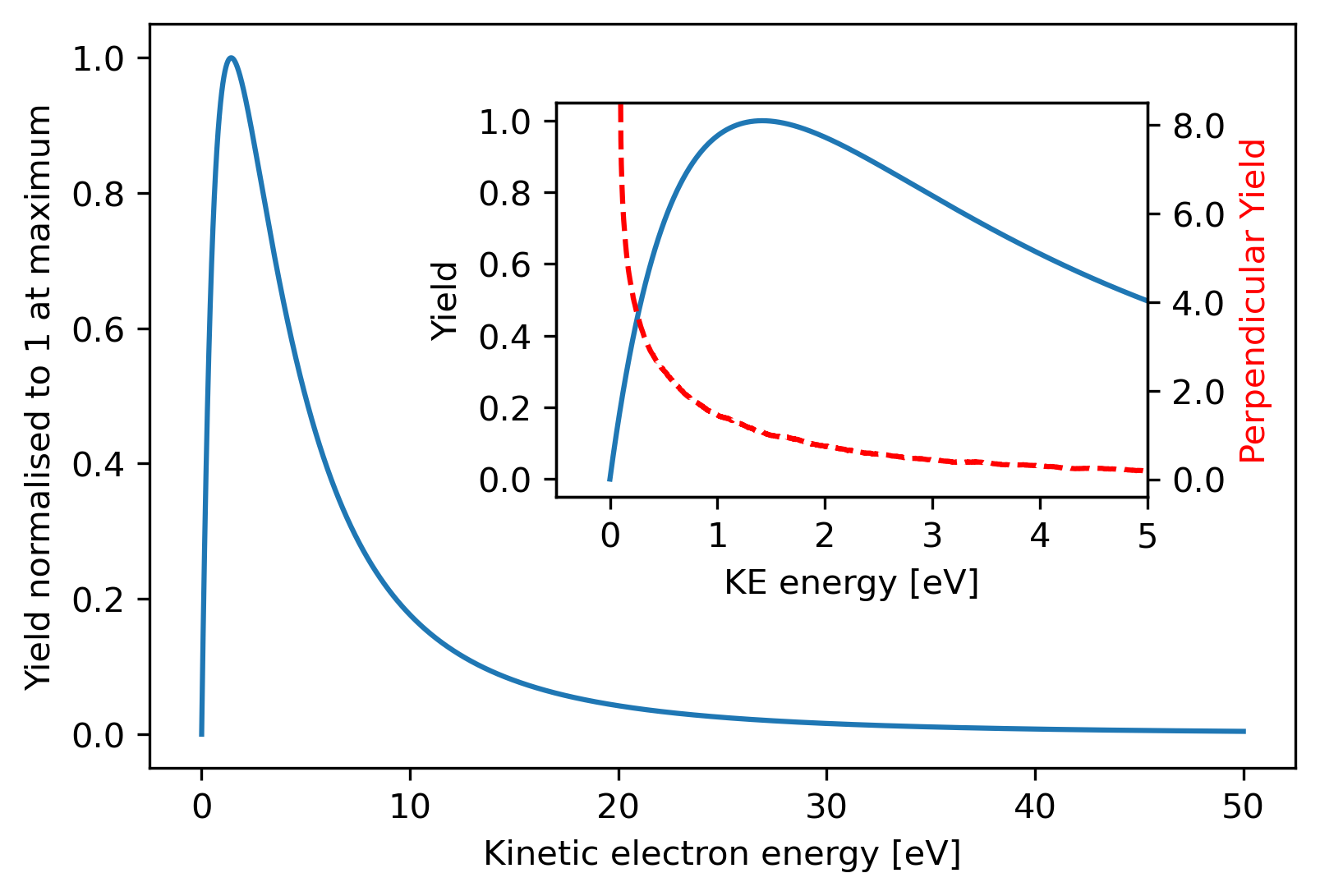}
\caption[KE Energy Distribution.]{\label{fig:KEenergyDistribution} The expected kinetic electron energy distribution for a gold surface with work function $\phi = 4.2$\,eV.  The peak is at $\phi/3 = 1.4$\,eV, the median is at $\phi$, and the mean is at $2\phi =8.4$\,eV. The dashed red curve in the insert shows the perpendicular energy distribution assuming a cosine emission law, with the right-hand axis values normalised to give the same overall number of electrons.}
\end{figure}

{\it KE spectrum.}
In the context of scanning electron microscopy, there is extensive literature on the properties of EIEE, with Ref.~\cite{Seiler1983} providing a comprehensive review. KE are emitted with a cosine angular distribution and there is no experimental evidence for the KE yield being different for a primary particle entering or exiting a surface. In addition, for primary electrons with energy greater than 100\,eV, the resulting EIEE energy distribution is expected to be independent of the primary energy \cite{Chung1974}. However, the shape of the energy distribution is dependent on an effective work function for kinetic emission, $\phi_{\textsc{ke}}$, and Ref.~\cite{yang2016} gives its form as:
\begin{equation}
	\label{eq: KEenergyDist} 
		\frac{\partial\delta}{\partial E_{k}} =  \frac{AE_{k}}{(E_{k}+\phi_{\textsc{ke}})^{4}} \,,
\end{equation}
where $A$ is a normalization factor, and $E_{k}$ is the kinetic electron energy. Both $A$ and $\phi_{\textsc{ke}}$ are dependent on any surface contamination~\cite{yang2016}. This form also seems to fit the predictions~\cite{taioli22} using a specific new model in \gf which deals with electron production from gold down to very low energies~\cite{sakata2016}, albeit with a relatively high value of $\phi_{\textsc{ke}}$.   For LPF the behaviour of the gold surfaces on TMs and surrounding electrodes/enclosures was examined in the context of the photoelectric discharge system. An average photoelectric work function of 4.2\,eV was found \cite{Armano2018a} which is lower than that of clean pristine gold, but is comparable to subsequent representative laboratory-based measurements \cite{wass2019,olatunde2020}. Taking Eq.~\ref{eq: KEenergyDist} and this work function gives the KE energy distribution shown in Fig.~\ref{fig:KEenergyDistribution}. As a consequence of the energy scale of this distribution, it is  expected that the transfer of secondary electrons between TM and electrode housing will be influenced by Volt-scale potentials. Actuation voltages used to control the TM position and the TM potential itself can reach this level. Therefore, depending on the number of secondary electrons emitted, an electrostatic dependence on the net TM charging rate is expected.  Given the parallel-plate geometry of the GRS, the fraction of secondary electrons affected by the TM potential will depend on the distribution of energies associated with the perpendicular component of the emitted electron velocities and this has been derived from Eq.~\ref{eq: KEenergyDist} using a cosine-weighted angular emission to give the dashed curve in the insert in Fig.~\ref{fig:KEenergyDistribution}.   

%%%%%%%%%%-----------------------------------------------------------------------%%%%%%%%%

\subsection{\label{SKE}Description of the Monte Carlo simulation tool}
\label{model:GRSSimulation}

The \gf simulation tool for calculating test-mass charging produced by cosmic rays is largely unchanged from previous work; hence only the principle of the simulation is described as further details are available in ~\cite{Araujo2005,Wass2005}

Within the simulation, cosmic ray particles of a given species are generated individually with an energy drawn from a distribution defined in the simulation setup. 
Each particle is generated on the surface of a generator sphere surrounding the spacecraft model with a random direction drawn from a cosine distribution about the surface normal. The result is an isotropic flux of particles within the simulation sphere. When simulating a differential flux of particles $J(E)$, the simulation time can be calculated from $T = N/A\int{J(E)}dEd\Omega$, where $A = \pi r^2$ is the cross-sectional area of the generator sphere, and $N$ is the number of particles simulated. This also allows us to allocate a mean time to the $N^{\mathrm{th}}$ simulated event.

The primary particle flux spectra used in the simulations have been updated to be as representative as possible of the flux experienced by LPF in orbit, and specifically for the times of the relevant experiments described in \cite{Armano2022}. Although protons and $^4$He are the dominant species, making up more than 96\% of cosmic ray particles in the energy regime relevant for charging, the simulations have also been carried out with $^2$H, $^3$He, $^{12}$C, and electrons, to include the next most abundant cosmic ray elements, and $^{56}$Fe nuclei to examine any effects associated with heavier elements. 

Proton flux data for the date of the analyzed experiment were taken from daily measurements of the cosmic ray flux by the Alpha Magnetic Spectrometer (AMS) on board the International Space Station~\cite{aguilar21b}. Helium fluxes were also derived from AMS flux data, albeit with lower temporal resolution~\cite{aguilar18}; these data are available until April 2017. For comparison with LPF  measurements from June 2017 the latest available data have been used.
For $^3$He AMS measurements~\cite{aguilar21} (of the $^3$He/$^4$He ratio) are used. PAMELA data are used for the galactic cosmic ray electron flux at solar minimum \cite{adriani11}, and for $^2$H nuclei~\cite{adriani13}. AMS data are used for $^{12}$C~\cite{aguilar17} and $^{56}$Fe~\cite{aguilar21c}. 

\begin{figure}[htb]
\centering
\includegraphics[scale=0.48, trim=2.5cm 3cm 4cm 14cm, clip]{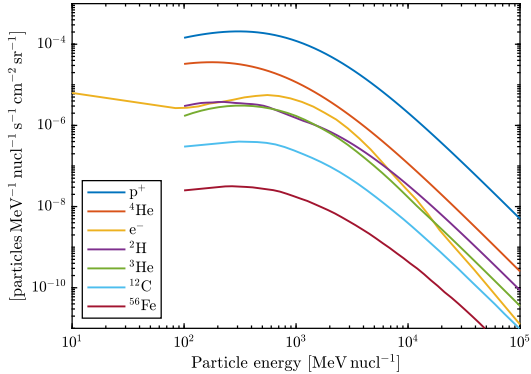}
\caption[KE Yield Curves.]{\label{fig:CR_spectra} The cosmic ray spectra used for this work.}
\end{figure}

Since contemporaneous data for heavier ions are not available, the uncertainty associated with these flux estimates is larger than for the dominant species. However, the abundance is low enough so that the contribution to the uncertainty in the total charge rate is small, as discussed later. 
In the case that measurement data do not cover the relevant energy range required as input for the simulation, an extrapolation was carried out so that all primary particles extend over the range 100\,MeV/nucleon to 1\,TeV/nucleon for hadrons. For electrons, a low-energy cutoff of 10\,MeV was chosen, just below the energy required to penetrate the 5.7\,g/cm$^2$ of material shielding the TMs. For proton and helium data from AMS, the lowest energy flux data is at around 500\,MeV/nucleon. To extrapolate to lower energies, an analytical description of the cosmic ray spectrum is adopted using the force-field approximation so that the flux $J$ as a function of energy, $E$, and the solar modulation parameter $\Phi$~\cite{Gleeson1968,Usoskin2011} can be written
\begin{equation}
    J(E, \Phi) = J_{\textsc{LIS}} \frac{E(E+2E_0)}{(E+\Phi)(E+\Phi+2E_0)}.
\end{equation}
Here $E_0 = mc^2$, with $m$ being the mass of the particle, and $J_{\textsc{LIS}}$ is the local interstellar cosmic ray spectrum. For best agreement with the data from AMS the parameterization of \cite{Burger2000} is chosen
for $J_{\textsc{LIS}}$:
\begin{equation}
    J_{\textsc{LIS}}(E)=\frac{1.9\times10^4\,P(E)^{-2.78}}{1+0.4866\,P(E)^{-2.51}}\, ,
\end{equation}
where $P(E) = \sqrt{E(E+2E_0)}$. Values of $\Phi=445$\,MV for protons and $\Phi=360$\,MV for alphas give best agreement with the data.
Figure~\ref{fig:CR_spectra} shows the energy distributions of primary particles used in the \gf simulations.

The total charge entering and leaving the TM in each event is tallied and recorded. In an extension from the previous work, the species of every particle that crosses a TM or opposing electrode housing surface is also recorded together with its location, direction, energy and event number. These particles serve as the primaries for a separate calculation of the KE emission process.

For each particle traversing a relevant boundary during a given event generated by the \gf simulation, the energy and angle-dependent KE yield curve (Fig.~\ref{fig:KEyieldCurves}) is used to determine the Poisson probability of KE emission.  Each electron is assigned a location based on the position where the primary particle traversed the surface and a time based on the simulation event number.
Each kinetic electron is assigned an energy sampled from the distribution shown in Fig.~\ref{fig:KEenergyDistribution}, assumed independent of primary particle energy and species.

An electron-tracking code developed for analysing the behaviour of the UV discharge system \cite{Armano2018a} is used to determine the influence of electrostatic fields on the low-energy component of the TM charging rate.
As in that work, the sensor is segmented into 24 parallel plate regions between the TM and housing corresponding to the equipotential surfaces on the electrode housing (18 electrodes plus 6 faces). An additional region of the housing is defined to account for corner regions without a parallel TM surface opposite and the grounded, exposed gaps around the electrodes.

To model the effect of electric fields within the sensor, fully representative voltages are applied to each electrode region.  This includes all time-varying actuation and injection voltages and the instantaneous TM potential contributed by charge accumulation and capacitively induced bias.

The ability of the emitted electron to traverse the gap between TM and housing is determined based on the instantaneous potential difference at the emission location. If the electron can traverse the gap between the TM and its surroundings, the change in the TM charge is recorded. This calculation is carried out for all KEs emitted by a primary event originating in the \gf simulation, with the electrode voltages evaluated at a time determined by the event number. 

In this simplified treatment, edge field effects are ignored. In reality, since electrons are emitted from the surface with some angular distribution, it is possible for them to escape the parallel field region and miss the TM even if they are energetic enough to traverse the local potential barrier. To account for this effect, an additional factor ($<1$) is applied to the probability of emission. This factor was determined for each surface in the sensor assuming a cosine angular distribution of emission and uniform spatial distribution. The fraction of KEs emitted from each region and directed toward or away from the TM was calculated by a simple Monte Carlo model.

The final result of the simulation is a timeline of charging events produced by primary cosmic ray interactions and secondary kinetic electron emission. Summing the two timelines, event by event, produces the final net charge timeline on each TM from which the net charging rate and charge-rate fluctuations are determined which can be compared with measurements made during the LPF mission.

%%%%%%%%%%-----------------------------------------------------------------------%%%%%%%%%

\subsection{A simple analytical model of the behaviour of the low-energy electron population}
In order to understand the features of the sensor design which influence the overall charging rate and its dependence on potential, a simple model is presented to  supplement the Monte Carlo simulation method in relation to the behaviour of the low-energy secondary electron population. 

The assumptions used in the model are:
\begin{itemize}
\item[(i)] The TM charging rate caused by the high energy cosmic ray processes is constant and unaffected by any applied potentials.  This overall rate is denoted by $H$.
\item[(ii)] The low-energy secondary electron emission rate per unit area, $R_s$, is the same from all surfaces, both TM and EH. 
\item[(iii)] When $V_{\textsc{tm}}=0$ the kinetic electrons will travel in straight lines away from their point of origin.  This means that all electrons leaving the TM will end up on the EH, whereas only a fraction, $\alpha$ of those leaving the EH will end up on the TM due to the gaps between the EH and the TM.
\item[(iv)] When $V_{\textsc{tm}} < 0$  the flow of kinetic electrons away from the TM will be unaffected as they still all leave the TM. However, those flowing from the EH will be repelled away from the TM and some fraction $\beta(V_{\textsc{tm}})$ will no longer reach it.  
\item[(v)] When $V_{\textsc{tm}} > 0$  the flow of kinetic electrons away from the TM will be reduced as some fraction $\beta(V_{\textsc{tm}})$ will no longer leave the TM. It is assumed that this fraction, $\beta$, has the same functional form as that for $V_{\textsc{tm}} < 0$. In addition the same fraction of electrons from the EH which would otherwise miss the TM (see point (iii) above) are now attracted to the TM.
\item[(vi)] In the regime of interest for the data acquired by LPF, $\beta(V_{\textsc{tm}})$ is determined by the perpendicular component of the energy distribution shown in Fig.~\ref{fig:KEenergyDistribution}. Its form depends on $\phi_{\textsc{ke}}$ and is best determined by numerical simulation.   
\end{itemize}

Hence, there are four free parameters in the model, $H$, $R_s$, $\alpha$, and $\phi_{\textsc{ke}}$. The model predictions are then
\begin{equation}
\begin{split}
    	\label{eq: qdot_neg} 
\dot{Q} (V_{\textsc{tm}}\le 0) =  H - \alpha A_h R_s \left(1+\beta(V_{\textsc{tm}}) \right) \\ + A_t R_s
\end{split}
\end{equation}
and 
\begin{equation}
\begin{split}
    	\label{eq: qdot_pos} 
\dot{Q} (V_{\textsc{tm}}\ge 0) =  H - \alpha A_h R_s \\ + A_t R_s \left(1-\beta(V_{\textsc{tm}}) \right) \\
		- \left(1-\alpha \right) A_h R_s \beta( V_{\textsc{tm}})\, ,
\end{split}
\end{equation}
where $A_h$ and $A_t$ are the surface areas of the electrode housing and TM. 

In routine operations of LPF there was a continuous sinusoidal TM potential of 0.6\,V applied at 100\,kHz.  Hence, in deriving results, from Eqs.~\ref{eq: qdot_neg} and \ref{eq: qdot_pos} a sine-weighted average must be performed around $V_{\textsc{tm}}$.

%%%%%%%%%%-----------------------------------------------------------------------%%%%%%%%%
%%%%%%%%%%-----------------------------------------------------------------------%%%%%%%%%

\section{Charging rate results}
\label{results:Q_rate}
This section describes the results of the test-mass charging models explained in Sec.~\ref{SKE} and compares with measurements made during the test campaigns described above.

\subsection{LPF measurements}
\label{LPF_data}

The variation in charging rate for TM\,1 and TM\,2 was measured on two occasions by LPF~\cite{Armano2022}. These are referred to by the dates on which the measurements were conducted.  The first, 2017-01-30, was a narrow-band data set with $V_{\textsc{tm}}$ between $\pm$0.3\,V.  The second, 2017-06-22, was over a wider band with $V_{\textsc{tm}}$ between $\pm$1\,V, together with two extra individual data points taken under different conditions.

\begin{figure}[htb]
\centering
\includegraphics[scale=0.48, trim=3cm 3cm 4cm 14cm, clip]{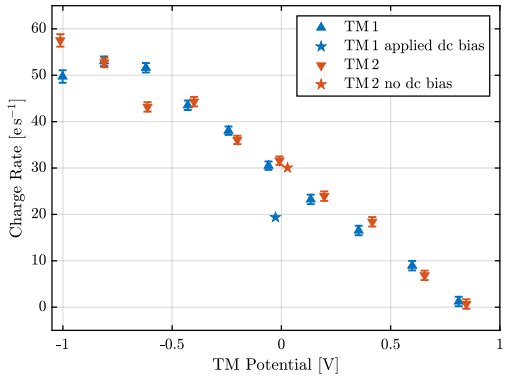}
\caption[Qrate vs TMv.]{\label{fig:Qrate_TMv} The variation in the TM charging rates observed with varying TM potential for the wide-band data set 2017-06-22.}
\end{figure}

Figure~\ref{fig:Qrate_TMv} shows the evaluation of the charging rate at each step of TM potential on TM\,1 and TM\,2 during the wide-band measurements of 2017-06-22 reproduced from Ref.~\cite{Armano2022}. The environmental charging rates of both TMs show similar, approximately linear dependencies on $V_{\textsc{tm}}$ with a slope of $\sim -30$\,e\,s$^{-1}$\,V$^{-1}$ and a charge rate at $V_{\textsc{tm}}$=0\,V of $28.3\pm0.8$\,e\,s$^{-1}$ on TM\,1 and $30.9\pm0.8$\,e\,s$^{-1}$ on TM\,2. The steep negative slope immediately confirms that low-energy electrons are playing a significant role and the symmetry either side of $V_{\textsc{tm}}$=0\,V implies that these are originating from both the TMs and from the surrounding surfaces.  The data show that for $V_{\textsc{tm}}\approx+1$\,V the negative low-energy electron charging balances the positive charging from the high-energy cosmic rays and the overall charging rate goes to zero. 
The additional data point for TM\,1 near 0\,V shows the result of the charge rate measurement made with a charged TM biased to $V_{\textsc{tm}}=0$\,V with dc voltages of 4.8\,V applied to the $y$ and $z$ electrodes. This measurement allows a comparison of the model predictions with measured data in a higher energy regime compared to probing the charge-rate measurement dependence on test mass potential alone.
Under these conditions with the overall TM\,1 potential at $-0.03$\,V the TM charging rate was 19.3\,e\,s$^{-1}$. At the same time, the rate measured on TM\,2 with no applied voltages and $V_{\textsc{tm}}=+0.03$\,V was 30.2\,e\,s$^{-1}$.

%%%%%%%%%%-----------------------------------------------------------------------%%%%%%%%%

\subsection{Simulation results: charging rate}
\label{results:simulation}

Using the combination of charging models described in Sec.~\ref{model:charging} 
the environmental charging rate in the electrostatic conditions described above can be estimated.

In Table~\ref{Tab:CR_rates} the predicted charging rates and noise contributions from the different cosmic ray species are given, as well as the KE contributions at $V_{\textsc{tm}}=0$\,V for nominal values of the parameters for the KE models. The results for TM\,1 are presented; for TM\,2 the results are consistent within the model uncertainties. The results are based on 1000\,s of simulation for all cosmic ray species. The total charging rate predicted by the model is higher than that measured during the 2017-06-22 experiment by 9.2$\pm$0.2\,e\,s$^{-1}$; this discrepancy is discussed further in the following sections. At $V_{\textsc{tm}}=0$\,V, the contribution of the total charge rate from kinetic electrons is small (12\% of the total) and positive. The overall charging rate from the high-energy component of the model is lower than that reported in Ref. \cite{Wass2005,  Araujo2005} for solar minimum conditions by around 10\,e\,s$^{-1}$ though the relative contributions to charging rate of protons, $^4$He and $^3$He are comparable. As was found in the previous work, heavier elements make a larger contribution to charging relative to their flux levels. Nonetheless, the overall contribution from each of the elements heavier than $^4$He including $^{56}$Fe is less than 1\,e\,s$^{-1}$.

\begin{table*}[htb]
\caption{\small Model predictions for TM\,1 charging rates and noise with $V_{\textsc{tm}}=0$\,V. Contributions for each cosmic ray species from the high-energy (HE) and kinetic electron (KE) model components are shown, along with the measured rate from data set 2017-06-22 of Ref.~\cite{Armano2022} and noise from Ref.~\cite{Armano2017a}.}
\centering
{\footnotesize
\setlength{\tabcolsep}{4pt}
\renewcommand{\arraystretch}{1.4}
\begin{tabular}{l | r@{}l | r | r| r | r@{}l | r@{}l | r@{}l}
\hline\hline
          \multicolumn{3}{l|}{Cosmic rays} &\multicolumn{3}{c|}{Charging rates}               & \multicolumn{5}{c}{Charging noise, $\lambda_{\mathrm{eff}}$}  \\
\hline
Species   & \multicolumn{2}{c|}{Flux} &\multicolumn{1}{c|}{HE}          & \multicolumn{1}{c|}{KE}  & \multicolumn{1}{c|}{Total} & \multicolumn{2}{c|}{HE} & \multicolumn{2}{c|}{KE} & \multicolumn{2}{c}{Total }  \\
          & \multicolumn{2}{c|}{[cm$^{-2}$\,s$^{-1}$]}&\multicolumn{1}{c|}{[e\,s$^{-1}$]} & \multicolumn{1}{c|}{[e\,s$^{-1}$]} &   \multicolumn{1}{c|}{[e\,s$^{-1}$]} & \multicolumn{2}{c|}{[e\,s$^{-1}$]} & \multicolumn{2}{c|}{[e\,s$^{-1}$]} & \multicolumn{2}{c}{[e\,s$^{-1}$]}\\
\hline
protons	  & 4&.28 &$26.6\pm0.6$  & $2.7\pm0.4$   & $29.3\pm0.7$     & 244&$\pm$6           & 146&$\pm$2      & 390&$\pm$8 \\
$^4$He	  & 0&.438 &$5.8\pm0.3$   & $0.9\pm0.2$    & $6.7\pm0.4$     & 89&$\pm$20           & 42&$\pm$3      & 131&$\pm$23  \\
$^3$He    & 0&.0507 &$0.7\pm0.1$   & $0.1\pm0.1$    & $0.8\pm0.1$     & 2.6&$\pm$0.2         & 2.5&$\pm$0.3    & 5.1&$\pm$0.5   \\
$^2$H	  & 0&.0671 &$0.4\pm0.1$   & $0.1\pm0.1$    & $0.5\pm0.1$     & 7.3&$\pm$0.7         & 3.4&$\pm$0.4    & 11&$\pm$1   \\
$^{12}$C	      & 0&.00742 &$0.2\pm0.1$   & $0.1\pm0.1$    & $0.3\pm0.1$     & 5.9&$\pm$0.9         & 2.2&$\pm$0.3    & 8&$\pm$1   \\
$^{56}$Fe        &  0&.000676&$0.2\pm0.2$   & $0.1\pm0.1$    & $0.3\pm0.2$     & 5&$\pm$5             & 2&$\pm$2        & 7&$\pm$6   \\
electrons & 0&.0554 &$-0.3\pm0.1$   & $-0.3\pm0.1$   & $-0.6\pm0.1$    & 21&$\pm$2            & 5.0&$\pm$0.3    & 26&$\pm$2   \\
\hline
Total     & \multicolumn{2}{c|}{---} &$33.6\pm0.6$  & $3.6\pm0.5$   & $37.5\pm0.8$     & 375&$\pm$33  & 202&$\pm$8  & 578&$\pm$42  \\
\hline
Measured  &  \multicolumn{2}{c|}{---} & \multicolumn{1}{c|}{---} &  \multicolumn{1}{c|}{---}  & \multicolumn{1}{c|}{$28.3\pm0.8$} & \multicolumn{2}{c|}{---}  & \multicolumn{2}{c|}{---}  & 1500&$\pm$180\footnote{Note that the cosmic ray flux at the time of the noise measurement was $\approx$ 20\% lower than during 2017-06-22 and so the value has been adjusted upwards to allow direct comparison.}  \\
\hline
\hline
\end{tabular}
}
\label{Tab:CR_rates}
\end{table*}

%%%%%%%%%%-----------------------------------------------------------------------%%%%%%%%%

\subsection{Charge noise}
\label{results:noise}

\begin{figure}[htb]
\centering
\includegraphics[scale=0.48, trim=3cm 3cm 4cm 14cm, clip]{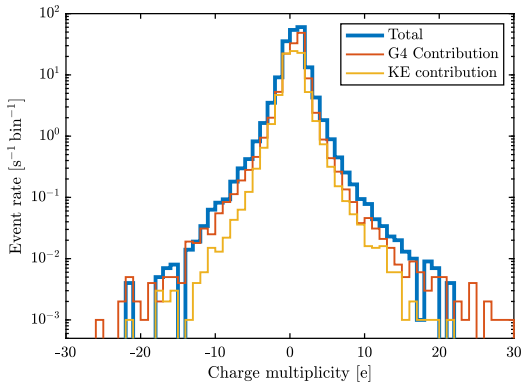}
\caption[Charge Multiplicity]{\label{fig:ChargeMultiplicity} A histogram of the multiplicity of charge deposits from the \gf and kinetic electron portions of the TM charging simulation. The multiplicity of all combined charging events is shown by the thick line. Note that this is the sum of the \gf and KE multiplicities event by event, not bin by bin in the histogram.}
\end{figure}

From Ref.~\cite{Araujo2005} the near balance of the rate of negative and positive charging events combined with the occasional high multiplicity event within the high-energy processes resulted in a predicted effective stochastic noise rate of $\lambda_{\mathrm{eff}}\sim$200--300\,e\,s$^{-1}$ in the absence of KE, with an increase to $\sim$400\,e\,s$^{-1}$ if the then estimated KE contribution was included.  Measurements of the stochastic noise rate made by LPF showed that the effective noise rate is even higher at $\lambda_{\mathrm{eff}}\sim$1100--1400\,e\,s$^{-1}$\footnote{Note these figures need to be adjusted for incoming cosmic-ray flux (and spectrum) and to compare with Table~\ref{Tab:CR_rates} for example they need to be increased by 20\%} despite the mean charging rate being close to the predictions~\cite{Armano2017a}. Table~\ref{Tab:CR_rates} shows the latest estimate for the KE noise contribution during the measurement of 2017-06-22.  Nominally, there are equal numbers of KE electrons per second per unit area being emitted from the EH and TM surfaces. However, there is 18\% more surface area on the EH than on the TM which might imply an overall reduction in the positive charging rate. The results in Table~\ref{Tab:CR_rates} show that the combined effect of some of the electron trajectories from the EH not hitting the TM, and some ac bias voltages being present even when $V_{\textsc{tm}}=0$ conspire to produce an additional small positive charging rate.  However, many more electrons are exchanged between TM and EH and these will contribute to the noise. In addition, some of these will be associated with high-multiplicity events. 
As shown by Fig.~\ref{fig:KEyieldCurves} the most efficient production of KE is from sub-keV electrons, sub-MeV protons and $\sim$MeV alpha particles and not from primary cosmic ray protons or helium nuclei.  Hence they will come from heavily degraded secondary particles of which there may be many per cosmic ray primary.
Figure~\ref{fig:ChargeMultiplicity} illustrates the multiplicity of charging events from 1000\,s of simulated data. As was demonstrated in previous work a broad range of charging event multiplicities is observed in the HE contribution to charging. Also shown is the multiplicity of KE charging events which shows net charge transfer as large as $\pm$22\,e due to KE in single events.   
The noise contribution from the high-energy component of the charging model is 375\,e\,s$^{-1}$, slightly higher than previous work. The additional contribution from KE is 202\,e\,s$^{-1}$, two times higher than the level calculated by the simple analytical model which does not take into account high multiplicity events. 

As seen in Table~\ref{Tab:CR_rates} the total noise predicted by the \gf simulations, coupled to the low-energy KE model, and following the approach of \cite{Araujo2005}, is only 40\% of the value actually measured (once incoming cosmic ray flux has been suitably adjusted from the in-situ radiation monitor (RM) count rate).  In principle, it would be easy to enhance the low-energy electron flux to explain the full noise budget.  However, as will be seen in the next section, this is constrained by requiring a fit to the observed dependence of charging rate on TM potential.  

There will be additional charge rate noise caused by any variations in the GCR flux, but these are expected to be small.  Inherent variability in the incoming cosmic ray fluxes during the LPF mission were monitored by its radiation monitor (RM)~\cite{Armano2018b,Armano2018d,Armano2019}. Fluctuations seen on short timescales ($f>10\,\mu$Hz) were dominated by RM counting statistics rather than real fluctuations~\cite{Armano2018b}. 
Between 1 and 10\,$\mu$Hz the RM singles count rate revealed an amplitude spectral density with $1/f$ shape.
Using the measured correlation coefficient between RM single count rate and TM charging rate~\cite{Armano2022} the corresponding charge-rate noise amplitude spectral density is $1.4\times\frac{0.1\,\mathrm{mHz}}{f}$\,e\,s$^{-1}\,\rm{Hz}^{-1/2}$. This is a small contribution (1\% of the noise power) to the measured value of 10\,e\,s$^{-1}\,\rm{Hz}^{-1/2}$ between 0.1 and 1\,mHz~\cite{Armano2017a}. 

%%%%%%%%%%-----------------------------------------------------------------------%%%%%%%%%

 \subsection{Electrostatic dependence of charging}
 \label{results:voltage_dependence}
 
\begin{figure}[htb]
\centering
\includegraphics[scale=0.48, trim=3cm 3cm 4cm 14cm, clip]{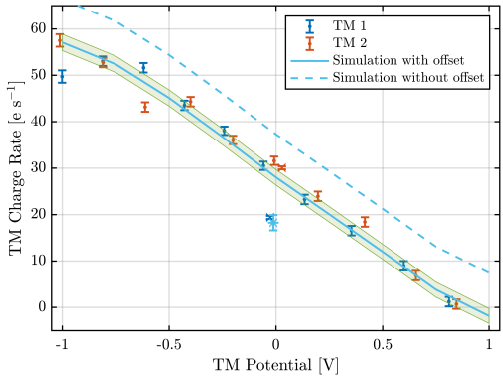}
\caption[Qrate vs TMv with sim]{\label{fig:Qrate_TMv_sim} The variation in the test mass charging rates observed with varying TM potential from the LPF 2017-06-22 data set, together with the simulated results assuming a low-energy kinetic electron population.}
\end{figure}

 {\it Monte Carlo Simulations}: The observed slope in Fig.~\ref{fig:Qrate_TMv} depends on the steeply falling part of the dashed curve in the insert of Fig.~\ref{fig:KEenergyDistribution} (applicable to the perpendicular emission component of velocity), and it can be seen that an applied bias of $\pm1$\,V can actually affect a relatively large fraction of the total KE population, with perpendicular energies below the bias potential. The value of the coefficient $Y_0$ in Eq.~\ref{ke_ang} determines the overall scale factor for the KE yield and larger values result in larger predicted slopes.  Figure~\ref{fig:Qrate_TMv_sim} shows the prediction of the model compared to the measured data using a set of parameters that give the best agreement with the TM\,1 data points. The shaded error region around the model prediction indicates the statistical uncertainties from the Monte Carlo simulation. In order to achieve the level of agreement shown, a constant offset has been applied to all simulation data. The uncorrected results are shown as a dashed curve. Based on results from the toy model of the KE charging process (see next subsection), an equal shift in the charging rate for all values of TM potential can be achieved through a change in the parameter $H$, equivalent to the charging rate in the absence of kinetic electrons, rather than a change in the KE emission model. A detailed discussion of the justification of the applied offset is provided in the following section. The best fit parameters for the electron yield model for both TM\,1 and TM\,2 data are $Y_0=1.15$, $E_0=700$\,eV and $s=1.64$. The value for the peak yield is at the lower end of yield estimates but still consistent with the literature. With these parameters, the best fit to the TM\,1 data is achieved with an offset of $-9.3$\,e\,s$^{-1}$ and $-8.5$\,e\,s$^{-1}$ for TM\,2. With the inclusion of the offset, the simulation model shows good agreement with the measured data ($\chi^2_{\mathrm{red}}=4.3$ and 5.4), including the additional measurement with applied bias in which the environmental charging rate is suppressed by around 10\,e\,s$^{-1}$ compared to the measurement at the same TM potential with no applied bias on TM\,2.
 
{\it Simple Model Results}: Equations~\ref{eq: qdot_neg} and \ref{eq: qdot_pos}, suitably averaged over the 100\,kHz bias, predict the variation in charging rate expected as a function of TM potential.  
 Allowing four free parameters in the simple analytical model does not result in a unique set of values for the best fit to the experimental data, and it is necessary to fix those values which are already reasonably well justified. This includes the effective KE work function which is taken to be similar to the photoelectric work function, with $\phi_{\textsc{ke}}=4.2$\,eV~\cite{Armano2018a,wass2019}.  Once this value has been set there is still too much degeneracy and the value of $\alpha$ is determined by geometry and by the KE angular distribution function, and should have a value just below 1. A preliminary scan of the residual $\chi^2$ for a range of values of $\alpha$ between 0.70 and 1.0, revealed a minimum at a value of 0.86, similar to the value from the Monte Carlo simulation, and this was then fixed.

\begin{figure*}[htb]
\centering
\includegraphics[width=0.46\textwidth]{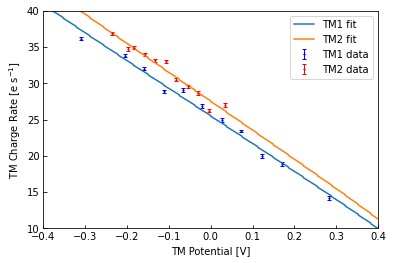}
\includegraphics[width=0.46\textwidth]{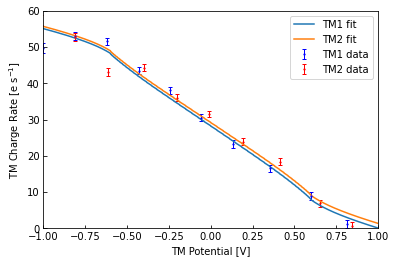}
\includegraphics[width=0.46\textwidth]{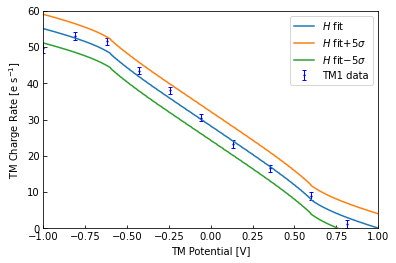}
\includegraphics[width=0.46\textwidth]{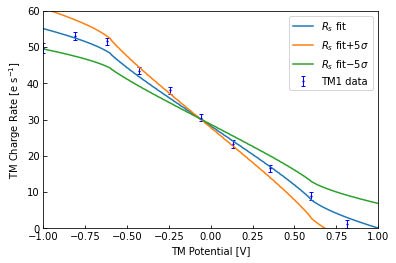}
\caption[Qrate vs TMv with model]{\label{fig:Qrate_TMv_model} The top panels show the variation in the TM charging rates observed with varying TM potential, together with the simple toy model results assuming a low-energy kinetic emission electron population. On the left are the fit results for the narrow range campaign, carried out on 2017-01-30, for TM1 and TM2.  On the right are the fits for the wide range campaign, carried out on 2017-06-22. The lower panels show the effects on the wide-band data fits of varying the toy model parameters, H and Rs, by $\pm5\sigma$ each, for the TM1 data.  The left lower panel shows the variation with H while the right lower panel shows the variation with Rs.  Similar variations are seen for the TM2 fits and for the narrow-band data.}
\end{figure*}

With the two remaining free parameters, $H$ and $R_s$, the simple model gives best-fit curves as shown in the upper panels of Fig.~\ref{fig:Qrate_TMv_model} for the two measurement campaigns.  The best fit values of the two free parameters and their uncertainties from the covariance matrices of the fit are shown in Table~\ref{Tab:model_params}.  The lower panels in Fig.~\ref{fig:Qrate_TMv_model} show how the charging rate curves change as the parameters are varied by $\pm5\sigma$.  From the plots it can be seen that $H$ moves the curve vertically as expected, whereas $R_s$ causes a rotation about a centre on the curve between $V_{\textsc{tm}}=-0.25$\,V and $V_{\textsc{tm}}=+0.25$\,V depending on the value of $\alpha$.  Altering the input cosmic ray flux will simultaneously change $H$ and $R_s$ causing a rotation about the point where the charging rate goes to zero.  
 
 The last row for each range in Table~\ref{Tab:model_params} gives the value of $R_s \times (A_h+A_t)$ which is the total number of low-energy kinetic emission electrons implicated in the population, $N_{\textsc{ke}}$.  However, it should be noted that these values are sensitive to $\alpha$, $\phi_{\textsc{ke}}$, the assumed form of $\beta(V_{\textsc{tm}})$ and to whether any other voltages are present, so care should be taken not to over-interpret the results from this simple model but more to use it to consolidate the understanding of the results from the full simulation. Comparing $N_{\textsc{ke}}$ with the overall KE charging rate and KE noise in Table~\ref{Tab:CR_rates} it can be seen that at $V_{\textsc{tm}}=0$ only a few percent of the electrons contribute to the charging rate and that there is a clear multiplicity enhancement in the KE noise contribution. Although the narrow range measurements require a statistically higher kinetic electron flux, implying a higher incoming cosmic ray flux, this is probably an artefact from having a shorter lever arm to fit over and if a fit is performed on the corresponding central points of the wide range data the values of $R_s$ become statistically consistent.

 In Table~\ref{Tab:model_params} results for a high value of $\phi_{\textsc{ke}}$ (15\,eV) are given.  This is motivated both by an observation in \cite{yang2016} that such a high value has been noted for Au after cleaning by sputtering, and by the recent work of \cite{taioli22} which shows a \gf simulation of the yield from Au which has a broad energy distribution which can be approximated to a function of the type shown in Equation~\ref{eq: KEenergyDist} with $\phi_{\textsc{ke}} \approx 15$\,eV.  A harder KE spectrum means fewer of the electrons will be affected by the applied bias, $V_{\textsc{tm}}$, and hence more are needed to obtain the observed dependence on $V_{\textsc{tm}}$. Adopting $\phi_{\textsc{ke}}=15$\,eV increases the overall number of KE electrons by 60\%, simultaneously increasing the associated KE charging noise contribution, shown in Table~\ref{Tab:CR_rates}, by the same proportion, thus partially closing the gap between model and observation in this respect.  However, even this increase in $\phi_{\textsc{ke}}$ is poorly motivated as the surfaces in LPF were large area and likely to be contaminated by exposure to air, which is in itself responsible for a lowered photoelectric work function~\cite{wass2019}, and a lowered value of $\phi_{\textsc{ke}}$ compared to a very clean surface~\cite{yang2016}.

\begin{table}[htb]
\caption{\small Parameters values for the simple model. Results for the two data sets are first given for a nominal work function of $\phi_{\textsc{ke}}=4.2$\,eV. Below that is a result for 2017-06-22 for a higher work function of $\phi=15$\,eV, which is a reasonable approximation to the KE energy distribution shown in Ref.~\cite{taioli22}.}
%\vspace{5pt}
\centering
{\footnotesize
\setlength{\tabcolsep}{4pt}
\renewcommand{\arraystretch}{1.4}
\begin{tabular}{l | r | r }
\hline\hline
Parameter & TM1 & TM2 \\
%&  &  \\
\hline
$\alpha$ & $0.86$	& $0.86$\\
$\phi_{\textsc{ke}}$ [eV] & $4.2$ & $4.2$ \\
\hline
2017-01-30 & & \\
$H$	[e\,s$^{-1}$]  & $27.9\pm 0.2$ & $30.1\pm 0.3$ \\
$R_s$ [e\,s$^{-1}$m$^{-2}$] & $4620\pm 171$	& $4840\pm 344$ \\
$N_{\textsc{ke}}$ & $128\pm5$    & $134\pm10$  \\
\hline
2017-06-22 & 	& \\
$H$	[e\,s$^{-1}$]  & $30.2\pm 0.8$ & $31.2\pm 0.9$ \\
$R_s$ [e\,s$^{-1}$m$^{-2}$] & $3790\pm 171$	& $3744\pm 195$ \\
$N_{\textsc{ke}}$ & $105\pm4$    & $104\pm5$  \\
\hline
\hline
$\phi_{\textsc{ke}}$ [eV] & $15.0$ & $15.0$ \\
\hline
2017-06-22 & 	& \\
$H$	[e\,s$^{-1}$]  & $30.7\pm 0.8$ & $31.6\pm 0.9$ \\
$R_s$ [e\,s$^{-1}$m$^{-2}$] & $6263\pm 294$	& $6190\pm 314$ \\
$N_{\textsc{ke}}$ & $174\pm8$    & $172\pm9$  \\
\hline
\hline
\end{tabular}
}
\label{Tab:model_params}
\end{table}

%%%%%%%%%%-----------------------------------------------------------------------%%%%%%%%%
%%%%%%%%%%-----------------------------------------------------------------------%%%%%%%%%

\section{Discussion}
\label{discussion}

The following section discusses the completeness and uncertainties in modelling of TM charging, paying particular attention to the need to apply an offset in the electrostatic dependence in the data compared to predictions. The charging rates and noise due to kinetic electrons shown in Table \ref{Tab:CR_rates} imply that the contribution of negative and positive charging currents are nearly balanced at a neutral TM potential. Increasing or decreasing the kinetic electron flux will therefore have only a minimal effect on the charging rate at $V_{\textsc{tm}}=0$\,V. Cosmic ray flux and the physics of charging by higher-energy particles are the main systematic uncertainties in estimating the charging rate in the absence of kinetic electrons.

\begin{figure}
    \centering
    \includegraphics[scale=0.48, trim=3cm 3cm 4cm 14cm, clip]{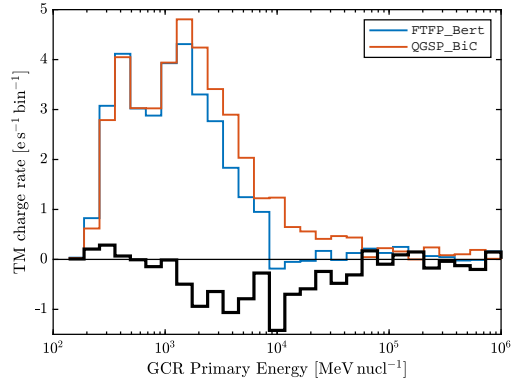}
    \caption[Charge Spectrogram]{\label{fig:ChargeSpec}Charging spectrogram---net charge deposited as a function of the energy of the primary particle for the event---for the two hadronic physics lists used in \gf. The black curve indicates the difference between the \texttt{QGSP_BiC} and \texttt{FTFP_Bert} models.}
    \label{fig:my_label}
\end{figure}

{\it Cosmic Ray Fluxes}: Test mass charging at all energy ranges depends on the cosmic ray flux through the spacecraft. Variations in this flux are not independent of particle energy, and the biggest variability comes in the energy range with the highest flux (100\,MeV--10\,GeV) and which produces most of the TM charging by both kinetic electrons and higher energy particles (Fig.~\ref{fig:ChargeSpec}). To a first approximation, therefore, the charging rate is proportional to the integral cosmic ray flux above 100\,MeV. The uncertainty on flux levels reported by AMS results in a $\pm2\%$ uncertainty in the particle flux which corresponds to an uncertainty in the charging rate of $<1$\,e\,s$^{-1}$ at $V_{\textsc{tm}}=0$\,V. 

Not all cosmic ray species have been included in the simulation; however, the six most abundant species have been analyzed. $^56$Fe was also simulated to check for effects from heavy ions that may contribute significantly despite low flux levels. Only protons and $^4$He contribute more than 1\,e\,s$^{-1}$ to the net charging rate. Missing cosmic ray species therefore account for an uncertainty of $<1$\,e\,s$^{-1}$ in the neutral TM charging rate.   

Variability of the incoming cosmic ray fluxes during the LPF mission has been observed by its radiation monitor~\cite{Armano2018b,Armano2018d,Armano2019} and by AMS~\cite{aguilar21b}. Using the data from the LPF radiation monitor taken during the charge rate measurements analyzed here, it is  concluded that there is no variation of the cosmic ray flux in the energy range relevant for charging above the $1\%$ level. Flux variations are therefore a negligible contribution to the measurement error.  

Systematic differences between the flux observed by AMS on the ISS and LPF at the L1 Lagrange point are possible but difficult to quantify. Measurements of the absolute cosmic ray flux at LPF have been performed~\cite{Armano2018b} but are susceptible to systematic uncertainties from particle tracking simulations in a similar way to the TM charging calculations. For this reason the AMS measurements for absolute flux levels can be relied upon and it is assumed that the additional flux uncertainty due to the separation of AMS and LPF is small. The long-term stability of the particle flux as observed on board LPF implies there were no flux changes propagating between L1 and Earth orbit during the period of measurement.

{\it Particle transport}: As described in Section \ref{model:G4_sims}, for the \gf part of the simulation there is a choice of physics models governing various energy ranges. The impact of higher energy physics models on the charging rate was investigated by comparing results with the \texttt{QGSP\_BiC} and  \texttt{FTFP\_Bert} hadronic physics lists. The total charging rate at V$_{\textsc{tm}}=0$\,V for the \texttt{QGSP\_BiC} models is 8\,e\,s$^{-1}$ higher than the values \texttt{FTFP_Bert} model presented in Table~\ref{Tab:CR_rates} and closer to the predictions of Refs.~\cite{Araujo2005, Wass2005}. The majority of this difference (7.4\,e\,s$^{-1}$) comes from the high-energy part of the simulation. Figure~\ref{fig:ChargeSpec} shows the TM charging spectrogram for the two simulations, confirming that the charging contribution for primary particles with energies in the range 1--50\,GeV contribute significantly less positive charging with the \texttt{FTFP\_Bert} models. Inelastic hadron interactions in the range up to 10\,GeV are governed by the Bertini and binary cascade models, with the latter producing more positive charging. Above 10\,GeV primary energy the \texttt{QGSP} and \texttt{FTFP} alogrithms are used. At the highest energies the difference between models is not significant but, at the lowest energies at which these models are applied, there is also significantly more positive charging in the \texttt{QGSP} model.

\begin{figure}
 \centering
 \includegraphics[scale=0.48, trim=3cm 3cm 4cm 14cm, clip]{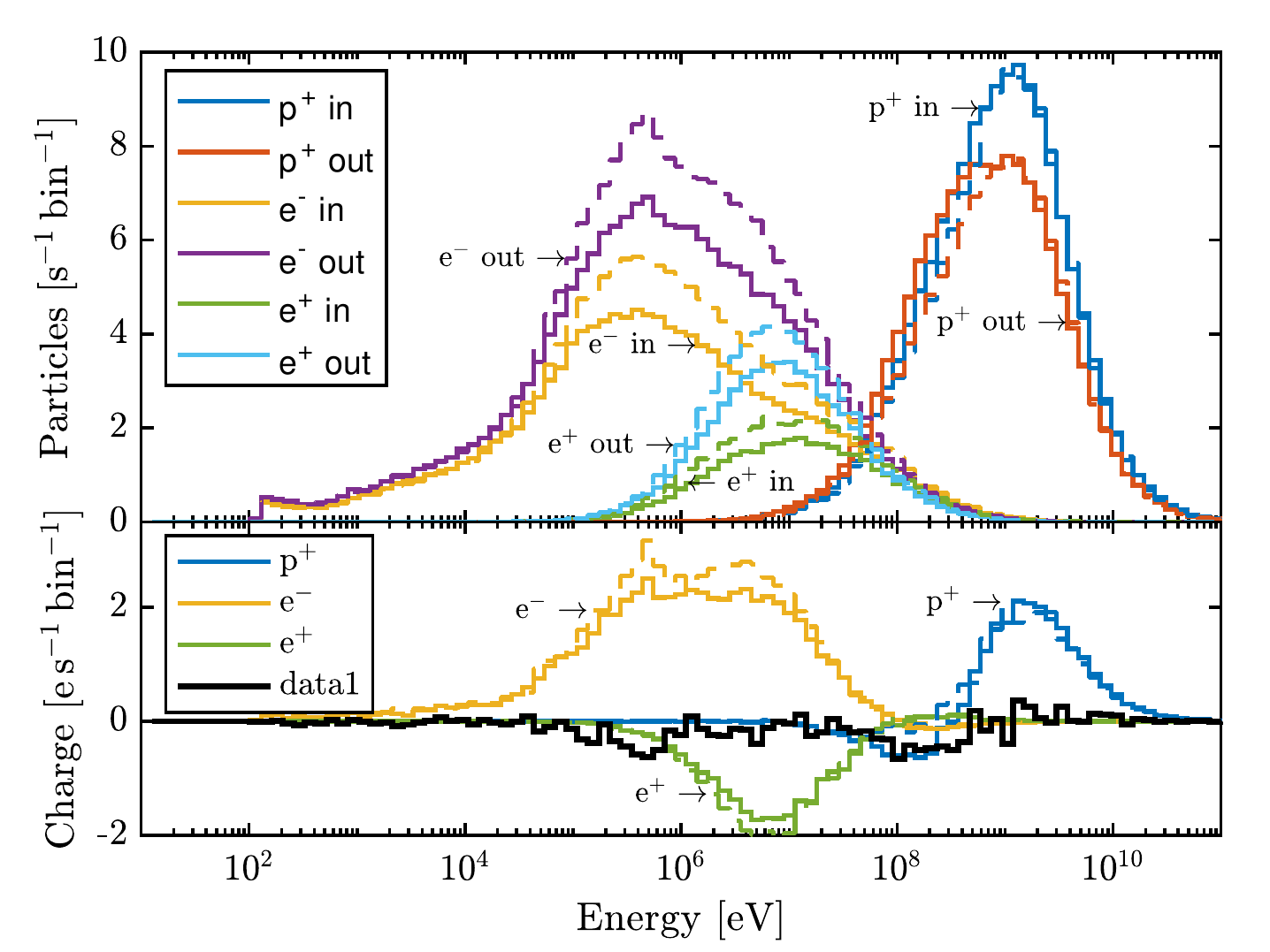}
 \caption[TM Particles In Out]{\label{fig:TMParticles} Top panel: energy distribution of particles entering and leaving the TM for a simulation using the \gf \texttt{FTFP_Bert} models (solid lines) and \texttt{QGSP_BiC} models (dashed lines). Lower panel: TM charging calculated from the above particle energy distributions for \texttt{FTFP_Bert} models (solid lines) and \texttt{QGSP_BiC} models (dashed lines). The black line shows the total difference in the distributions giving the net difference in charge rate between the two hadronic physics lists.}
 \end{figure}

Examining the distribution of particles entering and leaving the TM and electrode housing boundaries (upper panel of Fig.~\ref{fig:TMParticles}), it can be seen that there are differences in the electron and positron populations created by the two models at the TM boundary in the 30\,keV to 50\,MeV range with the \texttt{QGSP_BiC} model producing more of these secondaries. The \texttt{FTFP_Bert} model also produces a higher number of protons leaving the TM around 100\,MeV. When calculating the net charging of the TM by differencing the particles entering and leaving (lower panel of Fig.~\ref{fig:TMParticles}), it can be seen that the difference between physics models is smaller. The main contributors to the lower charge rate produced by \texttt{FTFP_Bert} can nonetheless be seen due to electrons at around 300--700\,keV and protons around 60--250\,MeV. 

Figure~\ref{fig:TMParticles} highlights the sensitivity of the predicted net charging rate to systematic differences in the high-energy physics models that produce secondary particles that result in significant TM charging. Although a detailed analysis of the differences between the models is beyond the scope of this work and difficult to analyse in a complex, integrated system such as the LPF spacecraft and GRS, assigning a systematic uncertainty consistent with the offset between simulation and measured results appears justified. 

{\it Unmodelled spacecraft structures}: Although the spacecraft geometry was updated from that used in Ref.~\cite{Wass2005} to be more representative of the as-flown structure for the TMs and their nearby environment, there was still some 20\% of mass missing in the simulation compared to the real spacecraft model.  To assess the effect of this an additional simulation run was carried out adding extra mass by increasing the density of materials used outside of the core sensor volume. The observed change in predicted charging rate was $<1$\,e\,s$^{-1}$. 

The remainder of the discussion below focuses on the uncertainties associated with the predictions of charge noise and the electrostatic dependence of the charging-rate by the model.  

{\it KE yield}: The parameters in the KE model are degenerate but constraining certain values, for example the effective work function that describes the energy distribution of KEs emitted from the sensor surfaces produces a set of parameters that is compatible with the data and KE yield values for electrons on gold quoted in the literature, e.g. Ref.~\cite{lin2005}. Analysing the variability of the parameters individually gives the following approximate uncertainties for TM\,1:  $Y_0 = 1.15\pm0.10$, $E_0 = 700\pm100$\,eV, $s=1.64\pm0.04$ and the offset applied to the model, $9.3\pm0.8$\,e\,s$^{-1}$. These variations in the yield parameters correspond to a change in the predicted charging rate at $V_{\textsc{tm}}=\pm1$\,V of $\pm1$\,e\,s$^{-1}$.

The KE yield parameters that fit the measured data are also correlated with the systematic uncertainty in the high-energy component of the charging process. If the high-energy component of the charging were systematically lower in reality than in the simulation, then it is likely that the number of primary particles capable of producing secondary emission would also be lower. This would result in a higher yield being required in order to fit the electrostatic charging dependence.   

{\it Transport of kinetic electrons in electric fields}: the electron cross-gap transport probability calculation in the KE model neglects edge field effects. A full electric field model based on applied electrode voltages would provide a more complete but more computationally intensive approach. The model assumes that electrons emitted from the EH on a trajectory that will miss the TM cannot contribute to charging. If this assumption is modified so that 5\% more electrons from each EH surface are able to reach the TM---a reasonable estimate of the uncertainty of this simple electrostatic approach---then the KE electron contribution to the net charging rate reduces to near 0\,e\,s$^{-1}$. 

{\it Kinetic electron energy distribution}: the electron energy distribution uses the model of \cite{yang2016} using an effective work function equivalent to the physical work function of the gold surfaces of around 4.2\,eV. On the other hand, Ref.~\cite{yang2016} describes measurements of the electron energy distribution of gold surfaces that are compatible with an effective work function of 13\,eV. Though these measurements were made on atomically clean gold surfaces this nonetheless suggests a bounding range for energy distributions. Shifting the work function parameter to higher values reduces the fraction of electrons that can be influenced by a TM potential in the range $\pm$1\,V. In order to remain consistent with the observations shown in Fig.~\ref{fig:Qrate_TMv_sim}, a higher yield value is needed. The model parameters that provide a good fit to the data in this case are: $Y_0$ = 2.3, at high-end of values found in the literature, but compatible with Ref.~\cite{lundgreen2020} for example, and with a larger offset of  $-13$\,e\,s$^{-1}$. This parameterization also produces slightly worse agreement with the measurement at $V_{\textsc{tm}}$=0\,V and applied biases of 4.8\,V predicting a charge rate of $16\pm2$\,e\,s$^{-1}$ compared to the measured $19.4\pm0.3$\,e\,s$^{-1}$ and $19\pm2$\,e\,s$^{-1}$ predicted with a work function of 4.2\,eV. One significant advantage of this parameter set is a higher current shot noise with $\lambda_{\textrm{eff}}\approx1000$\,s$^{-1}$, significantly closer to the adjusted measured value of 1500\,s$^{-1}$. Requiring an electron energy distribution representative of pure gold is problematic however, as the surfaces in the LPF GRS would be typical of surfaces that were exposed to air for timescales of days to weeks through the instrument integration and test procedures.

\begin{figure}[htb]
 \centering
 \includegraphics[scale=0.48, trim=2.5cm 3cm 4cm 14cm, clip]{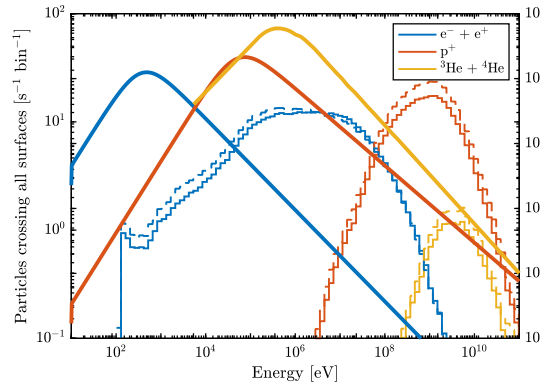}
 \caption[Secondary Particles]{\label{fig:SecondaryParticles} The distribution of secondary particles capable of producing kinetic electron emission crossing TM (solid lines) and EH surfaces (dashed lines). The smooth lines indicate the energy-dependent yields at normal incidence (right-hand axis) for the KE yield.}
 \end{figure}

{\it \gf ionisation production cuts}: As mentioned earlier there are two particle production energy cuts which limit the amount of secondary ionisation.  These are 100\,eV for ionisation produced by primary electrons, and 790\,eV for ionisation produced by primary protons. The effect of the production cut for primary electrons is seen as an abrupt cut-off at 100\,eV in Fig.~\ref{fig:SecondaryParticles}.  A small feature at 790\,eV is also evident. Hence, there are missing electrons within the simulation.  In addition if these secondary ionisation electrons have sufficient energy they could also add further KE electrons to the very-low energy population.  Adding to the KE population would steepen the slope of the charging rate dependency on TM potential and there would need to be a corresponding decrease in the overall KE yield parameter to maintain a good fit.  Hence to first order there would be little real effect from an extra KE source.  However, adding higher energy secondary ionisation electrons between 50 and 100\,eV would add to the charging rate noise without significantly affecting the slope of the dependence on potential and, moreover, could change the overall charging rate at $V_{\textsc{tm}}=0$\,V.  

Estimating the number of missing electron induced secondary ionisation electrons is helped by the clear identification of the population in Fig.~\ref{fig:SecondaryParticles} which increases with decreasing energy below $\sim$500\,eV.  Assuming this distribution continues to rise down to 10\,eV, as shown in Ref.~\cite{sakata2016}, would add a few 10s of electrons to the overall population. Proton-induced ionisation in gold has recently been studied using \gf in Ref.~\cite{rajabpour22}.  The secondary ionisation yield rises slowly down to 790\,eV, below which it then decreases to a minimum at around 200\,eV, an order of magnitude lower than the peak value. Hence, assuming the `bump' in the electron energy distribution in Fig.~\ref{fig:SecondaryParticles} above 790\,eV represents the peak of the distribution, the number of missing secondary electrons is conservatively likely to be similar to those from electron ionisation. The two together thus represent a small contribution to the overall electron numbers being ejected from the TM and EH surfaces and therefore the charging current shot noise.

{\it Unmodelled KE primaries}: KE yield measurements in the literature only exist for protons, electrons, and alpha particles. These make up $\sim$85\% of the particles crossing TM and EH boundaries generated by \gf. Although the yield from the other 15\% is unknown, it is likely that these particles will create KEs at some level leading to an underestimate of the KE population in the simulation (although the fitting procedure would compensate by adjusting the overall yield parameter). A conservative estimate would be to increase the number of KEs in the simulation by 15\% which would lead to a change in charging rate at the extremes of $V_{\textsc{tm}}$ of 2--3\,e\,s$^{-1}$.

%%%%%%%%%%-----------------------------------------------------------------------%%%%%%%%%
%%%%%%%%%%-----------------------------------------------------------------------%%%%%%%%%

\section{Conclusion}
\label{Discussion2}

A model has been presented that predicts the TM charging rate due to the cosmic ray environment of LPF. By tracking particles produced down to eV energies, the model explains the dependence of the charging rate of the TMs on the Volt-scale potential differences in the GRS with good accuracy. The ability to model these charging processes accurately in an arbitrary high-energy particle environment and electrostatic field configuration has applications in the development of future space instrumentation sensitive to KE emission.

The results of the model confirm that a significant fraction of the net charging current on the LPF test masses was made up of a low-energy population ($\sim$ eV) of electrons produced by electron- and ion-induced kinetic emission from the TM and surrounding sensor metallic surfaces. Constraints are placed on the the emission properties, yield and emitted energy distribution, that are consistent with previously measured values from gold although some degeneracy remains. A systematic discrepancy exists between the results of the charging model predictions and the observed data not related to the low-energy electron modelling. The most likely source of the uncertainty is in the modelling of the high-energy cosmic ray interactions with the spacecraft. 

The model predicts the stochastic properties of the charge build up but even with the inclusion of KE emission, the charge noise predicted is lower than measured in orbit. There is tension in the parameterization of the model between adding sufficient KEs to explain the observed noise behavior and the fitting of the electrostatic dependence of charging with realistic surface properties. Assuming a KE energy distribution consistent with a realistic gold work function of 4.2\,eV, 
an EIEE yield of 1.15 is required to explain the electrostatic charging dependence but the resulting charging current shot noise makes up only only 40\% of that measured in-orbit.  Allowing the gold work function to increase towards 15\,eV, with a yield of 2.3 would account for around 70\% of the noise power although this seems physically unlikely for a realistic gold surface. 

Uncertainties in the model have been extensively explored but there seems no clear route to breaking this tension and resolving the inconsistency between the predicted and measured charging rate. During the LPF mission, methods have been demonstrated to suppress charge-related noise effects to well below the acceleration noise budget for LISA and therefore high-precision knowledge of the charge noise is likely not required for mission success. If the effect were more critical to other future missions, an experimental test campaign to measure KE emission from representative gold surfaces could remove many of the uncertainties and degeneracies associated with the model.

%%%%%%%%%%-----------------------------------------------------------------------%%%%%%%%%
%%%%%%%%%%-----------------------------------------------------------------------%%%%%%%%%

\section*{Acknowledgements}
T. J. Sumner acknowledges support from the Leverhulme Trust (EM-2019-070\textbackslash4).

\bibliographystyle{unsrt2}

\bibliography{paper}

\end{document}